\documentclass[12pt]{article}
\usepackage{graphicx,psfrag}
\usepackage[nosort]{cite}

\newcommand \bra[1]{\left< {#1} \,\right\vert}
\newcommand \ket[1]{\left\vert\, {#1} \, \right>}

\newcommand{\bea}{\begin{eqnarray}}
\newcommand{\eea}{\end{eqnarray}}
\newcommand{\simgt}{\hbox{ \raise3pt\hbox to 0pt{$>$}\raise-3pt\hbox{$\sim$} }}
\newcommand{\simlt}{\hbox{ \raise3pt\hbox to 0pt{$<$}\raise-3pt\hbox{$\sim$} }}

\newcommand{\clfn}{\setcounter{footnote}{0}}

\begin{document}
\begin{titlepage}
\title{\bf \Large
\vspace{28mm}
QCD potential as a ``Coulomb--plus--linear'' potential\vspace{2cm}}
\author{Y.~Sumino
\\ \\ \\ Department of Physics, Tohoku University\\
Sendai, 980-8578 Japan
}
\date{}
\maketitle
\thispagestyle{empty}
\vspace{-4.5truein}
\begin{flushright}
{\bf TU--682}\\
{\bf March 2002}
\end{flushright}
\vspace{4.5truein}
\begin{abstract}
\noindent
{\small
We show analytically that the QCD potential can be expressed, up to an
${\cal O}(\Lambda_{\rm QCD}^3 r^2)$ uncertainty, as
the sum of a ``Coulomb'' potential (with log corrections at
short distances) and a linear potential,
within an approximation based on perturbative expansion in
$\alpha_S$ and the renormalon dominance picture.
The expansion of $V_{\rm QCD}(r)$ is truncated 
at ${\cal O}(\alpha_S^N)$ [$N \! = \! 6\pi/(\beta_0\alpha_S)$], 
where the term becomes
minimal according to the estimate by NLO renormalon,
and is studied for $N \gg 1$.
Analytic expressions for the linear potential are obtained
in some cases.
}
\end{abstract}
\vfil

\end{titlepage}

\section{Introduction}
\label{s1}

Analyses of the static QCD potential $V_{\rm QCD}(r)$
within perturbative QCD entered
a new phase when the cancellation of the leading order (LO) renormalons
between the QCD potential and the pole masses of quark and antiquark
was discovered \cite{renormalon}.
Convergence of the perturbative series improved dramatically and
much more accurate perturbative predictions became available.
Subsequently, several studies 
\cite{Sumino:2001eh,necco-sommer,Recksiegel:2001xq,Pineda,complat}
showed that perturbative
predictions for  $V_{\rm QCD}(r)$ agree well
with phonomenological potentials (determined from heavy quarkonium
spectroscopy) and lattice calculations of $V_{\rm QCD}(r)$, 
once the LO renormalon contained 
in the QCD potential is cancelled
(see also \cite{lee}).
In fact the agreement holds within
the perturbative uncertainty of ${\cal O}(\Lambda_{\rm QCD}^3 r^2)$ 
estimated from the residual next-to-leading order (NLO) renormalon \cite{al}.
Despite of different prescriptions used for cancelling the LO
renormalon, all these perturbative predictions were 
mutually consistent within the ${\cal O}(\Lambda_{\rm QCD}^3 r^2)$ 
uncertainty.\footnote{
This is true only
in the range of $r$ where the respective perturbative predictions are stable,
since all the perturbative predictions go out of control 
beyond certain distances.
}
These observations indicate validity of the renormalon dominance
picture for the QCD potential.

Empirically it is known that phenomenological potentials
and lattice computations of $V_{\rm QCD}(r)$ are both
approximated well by the sum of a Coulomb potential and a linear
potential in the range $r \simgt 0.5~{\rm GeV}^{-1}$ \cite{bali}.
The linear behavior at large distances
is consistent with the quark confinement picture.
For this reason, before the discovery of the renormalon cancellation,
it was often said that perturbative QCD is
unable to explain the ``Coulomb-plus-linear'' behavior of the QCD potential.

Once the cancellation of the LO renormalons is 
incorporated, the perturbative QCD potential gets steeper than
the Coulomb potential at large distances.
This feature can be understood, within perturbative QCD, 
as an effect of the running of the strong coupling constant 
\cite{bsv1,Sumino:2001eh,necco-sommer}.
%For instance, we may understand that the strong coupling constant
%defined from the interquark force (which does not contain LO renormalon)
%get larger in the infrared (IR), so that the interquark force becomes
%stronger than the Coulomb force at large distances.
%Another more microscopic description is as follows.
%The dominant components of the total energy of a color-singlet
%quark-antiquark pair 
%are the self-energies of quark and antiquark (which are positive),
%which include only contributions of gluons whose wavelengths are smaller
%than the interquark distance $r$.
%When $r$ gets larger, gluons with larger wavelengths contribute.
%Since these gluons have larger coupling to quark/antiquark, the self-energies
%increase rapidly, and so does the total energy.\footnote{
%If the coupling stays constant, the total energy is given by
%the Coulomb potential (up to an $r$-independent constant).
%}
On the other hand, it is not obvious whether 
the QCD potential is rendered to a ``Coulomb-plus-linear'' form by 
this effect.
The perturbative uncertainty due to the residual renormalon 
is of ${\cal O}(r^2)$, hence there is a possibility
that the ${\cal O}(r)$ term of the potential at 
$r \simlt \Lambda_{\rm QCD}^{-1}$ is predictable 
within perturbative QCD.
In this paper, by considering a
certain limit of a finite-order perturbative expansion of
$V_{\rm QCD}(r)$ based on the renormalon dominance picture, 
we show that indeed the potential can be decomposed
into a ``Coulomb-plus-linear'' form, up to
an ${\cal O}(\Lambda_{\rm QCD}^3 r^2)$ uncertainty.
Our prescription gives a prediction consistent with the previous
predictions 
\cite{Sumino:2001eh,necco-sommer,Recksiegel:2001xq,Pineda,complat}.

In Sec.~\ref{s2} we set up our conventions for our analysis.
Sec.~\ref{s3} presents an analysis in the large--$\beta_0$
approximation; Sec.~\ref{s4} presents an analysis based on
renormalization-group (RG), incorporating 1-, 2-, and 3-loop running of
the coupling constant.
Discussion and conclusions are given in Secs.~\ref{s5} and \ref{s6},
respectively.

\section{Perturbative QCD potential and renormalons}
\label{s2}

The static QCD potential is defined from an expectation value
of the Wilson loop as
\bea
V_{\rm QCD}(r) 
&=& 
- \lim_{T \to \infty} \frac{1}{iT} \,
\ln \frac{\bra{0} {\rm{Tr\, P}} \exp
\biggl[ i g_S \oint_{\cal P} dx^\mu \, A_\mu(x) \biggr]
\ket{0}}
{\bra{0} {\rm Tr} \, {\bf 1} \ket{0}}
\nonumber \\ 
&=& 
\int \frac{d^3\vec{q}}{(2\pi)^3} \, e^{i \vec{q} \cdot \vec{r}}
\, \biggl[
-4 \pi  C_F \, \frac{\alpha_V(q)}{q^2}
\biggr]
~~~;~~~~~
q = |\vec{q}|
,
\eea
where ${\cal P}$ is a rectangular loop of spatial extent $r$ and
time extent $T$.
The second line defines the $V$-scheme coupling contant,
$\alpha_V(q)$, in momentum space;
$C_F = 4/3$ is the second Casimir operator of the fundamental 
representation.
In perturbative QCD, $\alpha_V(q)$ is calculable in a series
expansion of the strong coupling constant:
\bea
\alpha_V(q) = \alpha_S \, \sum_{n=0}^{\infty} P_n(\ln (\mu/q) ) \,
\biggl( \frac{\alpha_S}{4\pi} \biggr)^n
= \alpha_S(q) \, \sum_{n=0}^{\infty} a_n \,
\biggl( \frac{\alpha_S(q)}{4\pi} \biggr)^n 
~~~;~~~
a_n = P_n(0) .
\label{alfV}
\eea
Here,
$P_n(\ell)$ denotes an $n$-th-degree polynomial of $\ell$.
In this paper, unless the argument is specified explicitly,
$\alpha_S \equiv \alpha_S(\mu)$ denotes the strong coupling constant
renormalized at the renormalization scale $\mu$,
defined in the $\overline{\rm MS}$ scheme.
The series expansion of $\alpha_S(q)$ in terms of $\alpha_S$
is determined by the RG equation
\bea
q^2 \, \frac{d}{dq^2} \, \alpha_S(q) = 
- \alpha_S(q) \sum_{n=0}^{\infty} \beta_n 
\biggl( \frac{\alpha_S(q)}{4\pi} \biggr)^{n+1},
\eea
where $\beta_n$ represents the $(n+1)$-loop coefficient of the
beta function.
Therefore, at each order of the expansion of $\alpha_V(q)$ in $\alpha_S$, 
the only part of the polynomial $P_n(\ln (\mu/q) )$ 
that is not determined by the RG equation is $a_n$.
The above equations fix our conventions.

It is known \cite{adm} that 
$a_n$ for $n \ge 3$ contain infrared (IR) divergences.
We will discuss this issue in Sec.~\ref{s5}, whereas
in Secs.~\ref{s3} and \ref{s4} 
we treat $a_n$ as finite constants.

According to the renormalon dominance picture,
the leading behavior of
the ${\cal O}(\alpha_S^{n+1})$ term of $V_{\rm QCD}(r)$
%, $V^{(n)}_{\rm QCD}(r)$, 
at large orders
is given by the LO renormalon contribution as
$V^{(n)}_{\rm QCD}(r) \sim
{ const.}\times n! \, (\beta_0 \alpha_S/(2\pi))^n \, n^{\delta/2}$, 
where $\delta = \beta_1/\beta_0^2$ \cite{Beneke99}.
In the computation of the heavy quarkonium spectrum,
the LO renormalon gets cancelled against the LO renormalons contained
in the quark and antiquark pole masses.
Considering this application, if we subtract the LO renormalon
contribution from $V^{(n)}_{\rm QCD}(r)$, its large-order behavior becomes 
${ const}.\times r^2 \, n! \, (\beta_0 \alpha_S/(6\pi))^n \, n^{3\delta/2}$
due to the NLO renormalon contribution. 
Then $V^{(n)}_{\rm QCD}(r)$ (after the LO renormalon is subtracted)
becomes minimal at
$n \approx N \equiv 6\pi/(\beta_0 \alpha_S)$
and its size scarcely changes for $N - \sqrt{N} \ll n \ll N + \sqrt{N}$.

In view of the usual property of asymptotic series,
we simply truncate the series expansion of the potential 
at the order where 
the term becomes minimal according to the renormalon dominance picture,
i.e.\ at ${\cal O}(\alpha_S^N)$:
\bea
V_N(r) \equiv [ V_{\rm QCD}(r) ]_N
= -4 \pi  C_F \,
\int \frac{d^3\vec{q}}{(2\pi)^3} \, \frac{e^{i \vec{q} \cdot \vec{r}}}{q^2}
\, [ {\alpha_V(q)} ]_N .
\eea
Here and hereafter, $[X]_N$ denotes the series expansion of $X$ in $\alpha_S$
truncated at  ${\cal O}(\alpha_S^N)$.
The purpose of this paper is to examine $V_N(r)$ for $N \gg 1$ while
keeping $\Lambda_{\overline{\rm MS}}$ \cite{LambdaMSbar}
finite, using certain 
estimates for the all order terms in Eq.~(\ref{alfV}).
The motivation for considering the large $N$ limit is that it corresponds
to the limits where the perturbative expansion becomes well-behaved 
(small expansion parameter)
%\footnote{
%This may be seen more vividly by considering the opposite limit:
%if the expansion parameter is large and the series diverges
%from low orders, it would not make sense to consider the 
%fixed-order perturbative expansion of $V_{\rm QCD}(r)$.
%} 
and where the estimate of $V^{(n)}_{\rm QCD}(r)$
by renormalon contribution becomes a better
approximation around $n \sim N$.
Note that large $N$ corresponds to small $\alpha_S$ and
large $\mu$ due to the relation between $N$ and $\alpha_S$.

Clearly, $V_N(r)$ cannot be
written as a ``Coulomb-plus-linear'' form for finite $N$,
since it is given as the Coulomb potential 
($-C_F \alpha_S /r$) times an
$(N \! - \! 1)$-th-degree polynomial of $\ln (\mu r)$, and therefore,
$V_N(r)\to 0$ as $r \to \infty$.
We will see, however, that $V_N(r)$ tends to 
a ``Coulomb-plus-linear'' potential (plus a quadratic potential)
 in the large $N$ limit.

\section{{\boldmath $V_N(r)$} in large--$\beta_0$ approximation}
\label{s3}

The large--$\beta_0$ approximation \cite{bb}
is an empirically successful method
for estimating higher-order corrections in perturbative QCD
calculations.
For $V_{\rm QCD}(r)$, this approximation corresponds to setting 
$a_n=(5\beta_0/3)^n$ and all $\beta_n=0$ except $\beta_0$.
(Therefore, it includes only the one-loop running of $\alpha_S(q)$.)
In this section, with these estimates of the all-order terms of 
$V_{\rm QCD}(r)$, we examine $V_N(r)$ for $N \gg 1$.
The reasons for examining the large--$\beta_0$ approximation are
as follows.
First, because this approximation leads to
the renormalon dominance picture; in fact, the renormalon dominance picture 
has often been discussed in this approximation.
Secondly, as stated in Sec.~\ref{s1}, the running of
the strong coupling constant makes the potential steeper at large distances
as compared to the Coulomb potential;
hence, we would like to see if the potential can be written as
a ``Coulomb-plus-linear'' potential when only
the one-loop running is incorporated as a simplest case.
We first present the results, discuss some properties, 
and then sketch how we derived our results.
\vspace{3mm}\\
{\it Results}\vspace{2mm}

We define 
$\widetilde{\Lambda} = e^{5/6} \,
\Lambda_{\overline{\rm MS}}^{\mbox{\scriptsize 1-loop}}$,
where 
$\Lambda_{\overline{\rm MS}}^{\mbox{\scriptsize 1-loop}}
= \mu \exp [-2\pi/(\beta_0 \alpha_S)]$.
In this section, we assume 
$
%\bea
e^{-5/6} \, \mu^{-1} = \widetilde{\Lambda}^{-1}e^{-N/3} \ll 
r \ll \widetilde{\Lambda}^{-1}e^{N/3}
%\label{limits}
%\eea
$
when taking various limits.
Note that, as $N \to \infty$,
the lower bound ($e^{-5/6} \, \mu^{-1}$) and the upper bound
($\widetilde{\Lambda}^{-1}e^{N/3}$) of $r$
go to 0 and $\infty$, respectively.

$V_N(r)$ for $N \gg 1$ within the large--$\beta_0$ approximation 
can be decomposed into four parts corresponding to
\{$r^{-1}$, $r^0$, $r^1$, $r^2$\} terms
(with logarithmic corrections in the $r^{-1}$ and $r^2$ terms):
\bea
&&
V_N^{(\beta_0)}(r) = \frac{4C_F}{\beta_0} \, \widetilde{\Lambda}
\,\, v(\widetilde{\Lambda}r,N) ,
\label{VNbeta0}
\\ &&
v(\rho,N)=v_C(\rho)+B(N)+C\rho+D(\rho,N)
+(\mbox{terms that vanish as $N\to\infty$}).
\label{vrhoN}
\eea
(i)``Coulomb'' part:
\bea
v_C(\rho)=-\frac{\pi}{\rho} + \frac{1}{\rho}
\int_0^\infty dx \, e^{-x} \, \arctan 
\biggl[ \frac{\pi/2}{\ln(\rho/x)} \biggr]
,
\eea
where $\arctan x \in [0,\pi)$.
The asymptotic forms are given by
\bea
\left\{
\begin{array}{ll}
v_C(\rho) \sim - \, \frac{\pi}{2\rho\ln(1/\rho)}, ~~~& \rho \to 0 \\
v_C(\rho) \sim - \, \frac{\pi}{\rho}, & \rho \to \infty \\
\end{array}
\right.
\eea
and both asymptotic forms are smoothly interpolated in the
intermediate region.

\noindent
(ii) constant part\footnote{
The ${\cal O}(1/N)$ and ${\cal O}(1/N^2)$ terms in eq.~(\ref{BN})
are irrelevant for $N \to \infty$.
We keep these terms in $B(N)$ for convenience in examining $V_N^{(\beta_0)}(r)$
at finite $N$; see Fig.~\ref{VN-CplusL} below.
}:
\bea
B(N)= - \int_0^\infty dt \, \frac{e^{-t}}{t} \,
\biggl[ \Bigl( 1 + \frac{3}{N}\, t\, \Bigr)^N - 1 \biggr]
- \ln 2 - \frac{9}{8N} + \frac{99}{64N^2} .
\label{BN}
\eea
The first term (integral) diverges rapidly for $N \to \infty$ as 
${\displaystyle - \, \frac{3}{2}\sqrt{\frac{2\pi}{N}}
\biggl( \frac{3}{e^{2/3}} \biggr)^N \,
\bigl[ \, 1 + {\cal O}(1/N) \bigr]}$.

\noindent
(iii) linear part:
\bea
C = \frac{\pi}{2} .
\eea

\noindent
(iv) quadratic part:
\bea
&&
D(\rho,N)=\rho^2 \, \biggl[ \frac{1}{12}  \ln N + d(\rho) \biggr],
\label{DrhoN}
\\ &&
d(\rho)= -
\int_0^\infty dx \, \,
\frac{e^{-x}-\Bigl[1-x+\frac{1}{2}x^2-\frac{1}{6}x^3 \, \theta(1-x) \Bigr]}
{x^4} \,\,
\frac{\ln (\rho/x)}{\ln^2 (\rho/x) + \pi^2/4}
\nonumber \\ && ~~~~~ ~~~
- \frac{1}{12}\, 
\biggl[ \ln \Bigl( \ln^2 \! \rho + \frac{\pi^2}{4} \Bigr) + 
\ln \frac{9}{2}
+ \gamma_E \biggr],
\label{drho}
\eea
where $\theta(x)$ is the unit step function and $\gamma_E = 0.5772...$
is the Euler constant.
The asymptotic forms of $d(\rho)$ are given by
\bea
\left\{
\begin{array}{ll}
\rule[-4mm]{0mm}{6mm}
d(\rho) \sim - \frac{1}{12}\, \Bigl[ 2 \ln \ln \frac{1}{\rho} + 
\ln \frac{9}{2} + \gamma_E \Bigr] ,~~~
& \rho \to 0 \\
d(\rho) \sim - \frac{1}{12}\, \Bigl[ 2 \ln \ln {\rho} + 
\ln \frac{9}{2} + \gamma_E \Bigr] , & \rho \to \infty \\
\end{array}
\right.
\eea
and in the intermediate region both asymptotic forms are smoothly interpolated.
\medbreak

Although the constant part of $V_N^{(\beta_0)}(r)$ 
diverges rapidly as $N \to \infty$,
the divergence can be absorbed into the quark masses
in the computation of the heavy quarkonium spectrum.
Therefore, in our analysis, 
we will not be concerned with the constant part of the potential
but only with the $r$-dependent terms.

The quadratic part of $V_N^{(\beta_0)}(r)$  diverges slowly as
$\widetilde{\Lambda}^3 r^2 \ln N  \sim 
\widetilde{\Lambda}^3 r^2 \ln \ln (\mu/\widetilde{\Lambda})$.\footnote{
Within the potential-NRQCD framework, this 
divergence or scale-dependence can be absorbed
into the ${\cal O}(r^2)$ term of
a non-local gluon condensate in the operator product expansion \cite{bpsv}.
}
We may consider
this feature to be a characteristic property of
renormalons for the following reasons.
(1) If the series expansion of
$m_{\rm pole}(m_{\overline{\rm MS}},\alpha_S)$ or $V_{\rm QCD}(r)$
is truncated at the order corresponding to the
minimal term of the LO renormalon contribution,
i.e.\ $N' = 2\pi/(\beta_0 \alpha_S)$, 
$[m_{\rm pole}]_{N'}$ or $[V_{\rm QCD}(r)]_{N'}$ diverges as 
$\widetilde{\Lambda} \ln N'$
(within the large-$\beta_0$ approximation).
We may compare $\widetilde{\Lambda} \ln N'$ with the usual interpretation 
that $m_{\rm pole}$ and $V_{\rm QCD}(r)$
contain ${\cal O}(\widetilde{\Lambda})$ perturbative uncertainties
due to the LO renormalons.
(2) We have checked that even if we incorporate the effect of the 
two-loop running,
i.e.\ even if we set $\beta_1 \ne 0$, 
the quadratic part of $V_N(r)$ still diverges as 
$\widetilde{\Lambda}^3 r^2 \ln N$.
Therefore, we interpret that the quadratic part of $V_N^{(\beta_0)}(r)$ 
represents an
${\cal O}(\widetilde{\Lambda}^3 r^2)$ uncertainty,
following the standard interpretation on the perturbative
uncertainty induced by the NLO renormalon.
\begin{figure}
\begin{center}
\psfrag{rho=Lambda*r}{\hspace{10mm}$\rho = \widetilde{\Lambda}\, r$} 
\psfrag{vc(rho)+C*rho}{$v_C(\rho)+C\rho$} 
\psfrag{lefttitle}
{\hspace{1mm}$v(\rho,N)\! - \! B(N)$ and $v_C(\rho) \! + \! C\rho$} 
\psfrag{N=10}{$N=10$}
\psfrag{N=30}{\hspace{-2mm}$\! \! N=30$}
\psfrag{N=100}{\hspace{-2mm}$\! \! N=100$}
\psfrag{r=1/Lambda}{$r=1/\widetilde{\Lambda}$}
\psfrag{r=1/LambdaMS}
{$r=1/\Lambda_{\overline{\rm MS}}^{\mbox{\scriptsize 1-loop}}$}
\includegraphics[width=10cm]{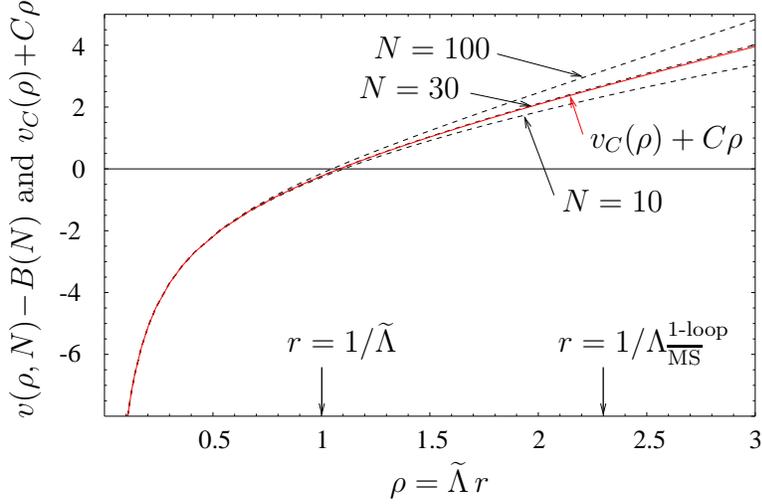}
\end{center}
\vspace*{-.5cm}
\caption{\small
Truncated potential after the constant term is subtracted,
$v(\rho,N)-B(N)$, (dashed black) vs.\ $\rho$ for 
$N=10$, 30, 100.
``Coulomb-plus-linear'' potential, 
$v_C(\rho)+C\rho$, (solid red)
is also plotted, which is hardly distinguishable from the 
$N=30$ curve.
\label{VN-CplusL}}
\end{figure}
In this respect, we note that the dependence of $V_N^{(\beta_0)}(r)$ 
on $N$ is mild; 
for instance, as shown in Fig.~\ref{VN-CplusL},
the variation of $V_N^{(\beta_0)}(r)$ is small
(after the constant part is subtracted) 
in the range $r \simlt \widetilde{\Lambda}^{-1}$ as we vary $N$ from
10 to 100; 
it corresponds to a variation of 
$\mu/\Lambda_{\overline{\rm MS}}^{\mbox{\scriptsize 1-loop}}$
from 30 to $3\times 10^{14}$.

The ``Coulomb'' part and the linear part are finite
as $N \to \infty$.
In Fig.~\ref{VN-CplusL}, we see that
$V_N^{(\beta_0)}(r)$ is approximated fairly well by the sum of the ``Coulomb''
part and the linear part (up to an $r$-independent constant) in the region 
$r \simlt \widetilde{\Lambda}^{-1}$ when we vary $N$ between
10 and 100.
Moreover, as long as $\frac{1}{12}\ln N \simlt {\cal O}(1)$, the 
difference between $V_N^{(\beta_0)}(r)$ and the ``Coulomb-plus-linear''
potential remains at or below ${\cal O}(\widetilde{\Lambda}^3 r^2)$
in the entire range of $r$.
\vspace{3mm}\\
{\it Outline of derivation}\vspace{2mm}

Let us write $L = \frac{\beta_0\alpha_S}{2\pi} 
\ln \Bigl(\frac{\mu e^{5/6}}{q}\Bigr)
= 1 + \frac{3}{N}\ln \Bigl(\frac{\widetilde{\Lambda}}{q}\Bigr)$.
$V_N^{(\beta_0)}(r)$ is defined as the Fourier transform of
$(4\pi C_F \alpha_S/q^2)\sum_{n=0}^{N-1}L^n
=(4\pi C_F \alpha_S/q^2) (1-L^N)/(1-L)$.
After integration over angular variables, 
it is given by
\bea
V_N^{(\beta_0)}(r) =
- \frac{2C_F\alpha_S}{\pi} \int_0^\infty dq \,
\frac{\sin (q r)}{q r} \, \frac{1-L^N}{1-L}
= \frac{4C_F}{\beta_0} \Bigl[
v_1(\widetilde{\Lambda}r) + v_2(\widetilde{\Lambda}r,N) \Bigr],
\label{orig}
\eea
where we separated the integral into two parts after deforming the integral
contour slightly:
\bea
&&
v_1(\rho) = {\rm Im} \int_0^\infty\! dk \,
\frac{e^{i k \rho}}{k\rho} \, \frac{1}{\ln (1/k) - i\epsilon}
= - \frac{1}{\rho} \, {\rm Im} \int_0^\infty\! dx \,
e^{-x} \, \ln \biggl[ \ln \Bigl(\frac{\rho}{x} \Bigr)-\frac{i\pi}{2} \biggr],
~~~
\label{v1}
\\ &&
v_2(\rho,N) = - {\rm Im} \int_0^\infty dk \,
\frac{e^{i k \rho}}{k\rho} \, \frac{1}{\ln (1/k) - i\epsilon}
\biggl[ 1 + \frac{3}{N}\ln \Bigl( \frac{1}{k} \Bigr) \biggr]^N
\nonumber \\ && ~~~~~ ~~~~~ ~
= - \frac{\pi \cos \rho}{\rho} - {\rm Pr.}
\int_0^\infty \! dk \, \frac{\sin (k\rho)}{k\rho}
\, \frac{1}{\ln(1/k)}
\biggl[ 1 + \frac{3}{N}\ln \Bigl( \frac{1}{k} \Bigr) \biggr]^N
\label{v2}
.
\eea
$v_1$ and $v_2$ are defined by the first equalities of
(\ref{v1}) and (\ref{v2}), respectively.
Contributions from the pole at $k=1$ in $v_1$ and $v_2$
cancel, since the original integral (\ref{orig}) does not 
contain a pole at $q=\widetilde{\Lambda}$.
In the second equality of (\ref{v1}), we deformed the integral
contour into the upper half plane on the complex $k$-plane
and integrated by parts.
As for $v_2$, since 
$[ 1 + \frac{3}{N}\ln \Bigl( \frac{1}{k} \Bigr) ]^N \to 1/k^3$
as $N \to \infty$,
the constant ($\rho^0$) and quadratic ($\rho^2$) terms 
in the integral become IR divergent in this limit.
On the other hand, the negative power of $k$ induces the positive
power behavior of $\rho$, i.e.\ the linear and quadratic terms,
in $v_2$ in the large $N$ limit.
We define
\bea
v_2(\rho,N) = \frac{A}{\rho} + B(N) + C\rho + D(\rho,N) 
+(\mbox{terms that vanish as $N\to\infty$}),
\eea
where $D(\rho,N) = {\cal O}(\rho^2)$.
In computing $A$ and $C$,
it is convenient to first remove divergences by subtracting
appropriate constant and quadratic terms from $v_2$.
Let
\bea
&&
\tilde{v}_2(\rho,N,k_0) 
\equiv - \frac{\pi \cos \rho}{\rho} - {\rm Pr.}
\int_0^\infty \! dk \, \biggl[ \frac{\sin (k\rho)}{k\rho}
- \theta(k_0-k) \biggl\{  1 - \frac{1}{6} (k\rho)^2 \biggr\}
\biggr]
\nonumber \\ && 
~~~~~~~~~~~~~~~~~~~~~~~~~~~~~~~~~~~~~~~~~~~~~~~~~~~~~~~~~~~~~~~
\times \, \frac{1}{\ln(1/k)}
\biggl[ 1 + \frac{3}{N}\ln \Bigl( \frac{1}{k} \Bigr) \biggr]^N
,
\label{tildev2}
\eea
where $k_0=2$ is an IR cutoff.
Now we may send $N \to \infty$ before integration over $k$.
$\tilde{v}_2(\rho,\infty,k_0) $ is finite for $0 < \rho < \infty$
and differs from ${v}_2(\rho,N)$
only by constant ($\rho^0$) and quadratic ($\rho^2$) terms,
apart from terms that vanish as $N \to \infty$.
One can show that the $\rho^{-1}$ and $\rho^1$ terms
stem only from the first term of (\ref{tildev2}):
\bea
&&
A = \lim_{\rho \to 0} \, \rho \, \tilde{v}_2(\rho,\infty,k_0) = - \pi ,
\\ &&
C = 
\lim_{\rho \to 0} \frac{1}{2}
\frac{\partial^2}{\partial \rho^2} [ \rho \, \tilde{v}_2(\rho,\infty,k_0) ]
= \frac{\pi}{2} 
.
\eea
The constant and quadratic terms can be calculated directly from $v_2$:
\bea
&&
B(N) = 
\lim_{\rho \to 0}\frac{\partial}{\partial \rho} [ \rho \, v_2(\rho,N) ] 
\nonumber
\\ && ~~~~~ ~~~
= - \int_{\epsilon}^\infty dt \, \frac{e^{-t}}{t} \,
 \Bigl( 1 + \frac{3}{N}\, t\, \Bigr)^N 
- \int_{-\infty}^{-\epsilon} dt \, \frac{e^{2t}}{t} \,
 \Bigl( 1-  \frac{9}{2N}\, t^2 + \cdots \, \Bigr) 
, 
\label{Binteg}
\\ &&
D(\rho,N) = 
v_2(\rho,N) - \biggl[ \frac{A}{\rho} + B(N) + C\rho \biggr]
\label{defD}
.
\eea
In the second equality of
(\ref{Binteg}) we set $t=\ln(1/k)$ and expanded the integrand in
$1/N$ in the region $t < 0$.
It is then straightforward to obtain (\ref{BN}).
One may separate a divergent part as $N\to\infty$ from
(\ref{defD}) in a similar manner.
As for the finite part ($N$-independent part), we
deform the integral contour into the upper half $k$-plane
to obtain (\ref{DrhoN}),(\ref{drho}).
Finally $v_C(\rho)$ is given by the sum of $A/\rho$ and
$v_1(\rho)$.\footnote{
Since the leading behavior of $V_N^{(\beta_0)}(r)$ as
$r\to 0$ is $const./(r \ln r)$ as determined by 
the 1-loop RG equation, the $A/\rho$
term of $v_2$ must be cancelled by the $1/\rho$
term contained in $v_1$.
}
The asymptotic forms of $v_C(\rho)$
 and $d(\rho)$ are obtained by expanding the
integrands in $\ln x$.

We made a cross check of our results by comparing  
$v(\rho,N)$ and 
$v_C(\rho)+B(N)+C\rho+D(\rho,N)$ numerically
for $3 \leq N \leq 1000$,
after subtracting the divergent terms from both.
The difference diminishes swiftly with $N$.

\section{{\boldmath $V_N(r)$} with 1-, 2-, and 3-loop
running of \boldmath{$\alpha_S(q)$}}
\label{s4}
\clfn

In this section we examine $V_N(r)$ in three cases corresponding
to the following estimates of the all order terms of $V_{\rm QCD}(r)$:
\\
(a) [1-loop running]
$\beta_0$, $a_0$: exact values, $\beta_n = a_n =0$ ($n \geq 1$);
\\
(b) [2-loop running]
$\beta_0$, $\beta_1$, $a_0$, $a_1$: exact values, 
$\beta_n = a_n =0$ ($n \geq 2$);
\\
(c) [3-loop running]
$\beta_0$, $\beta_1$, $\beta_2$, $a_0$, $a_1$, $a_2$: 
exact values \cite{tvz,ps}, 
$\beta_n = a_n =0$ ($n \geq 3$).
\\
We assume $\beta_0,\, \beta_1, \, \beta_2, a_0, a_1, a_2 
\,({\rm exact}) > 0$.\footnote{
This is the case when the number of 
quark flavors is less than 6 and all the quarks are massless.
}
In the standard 1-, 2-, and 3-loop RG improvements of 
$V_{\rm QCD}(r)$, the same all-order terms as in the above cases are resummed; 
the difference of our treatment is that
the perturbative expansions are truncated at ${\cal O}(\alpha_S^N)$.
We note that the renormalon dominance picture is consistent with
the above estimates of higher-order terms, or more generally, with
the RG analysis \cite{Beneke99}.
All the results for the case (a) can be obtained if we replace 
$\widetilde{\Lambda}$ by
$\Lambda_{\overline{\rm MS}}^{\mbox{\scriptsize 1-loop}}$
in the results of the large--$\beta_0$ approximation 
in Sec.~\ref{s3}.

Similarly to the previous section, we
can decompose $V_N(r)$ into four parts:
\bea
V_N(r) = V_C(r) + {\cal B}(N) + {\cal C} \, r + {\cal D}(r,N)
+ (\mbox{terms that vanish as $N\to\infty$}) ,
\label{decgen}
\eea
%The ``Coulomb'' and linear parts have finite limits as
%$N \to \infty$, whereas the constant and quadratic parts are divergent:
where
\bea
&&
V_C(r) = - \frac{4\pi C_F}{\beta_0 r} 
- \frac{2C_F}{\pi} \, {\rm Im}
\int_{C_1}\! dq \, \frac{e^{iqr}}{qr} \, \alpha_V(q) 
,
\label{genform1}
\\ &&
{\cal B}(N) = \lim_{r\to 0} \, \frac{2C_F}{\pi} \,{\rm Re}
\int_{C_1} \! dq \, e^{iqr} \, 
\Bigl\{ \alpha_V(q) - [ \alpha_V(q) ]_N  \Bigr\},
\\ &&
{\cal C} = \frac{C_F}{2\pi i} \int_{C_2}\! dq \, q \, \alpha_V(q) ,
\label{genform3}
\\ &&
{\cal D}(r,N) = V_N(r) - [V_C(r) + {\cal B}(N) + {\cal C} \, r] .
\label{genform4}
\eea
\begin{figure}
\begin{center}
\begin{tabular}{cc}
\psfrag{C1}{$C_1$} 
\psfrag{q}{$q$} 
\psfrag{q*}{$q_*$} 
\psfrag{0}{\hspace{-1mm}\raise-1mm\hbox{$0$}}
\includegraphics[width=4.5cm]{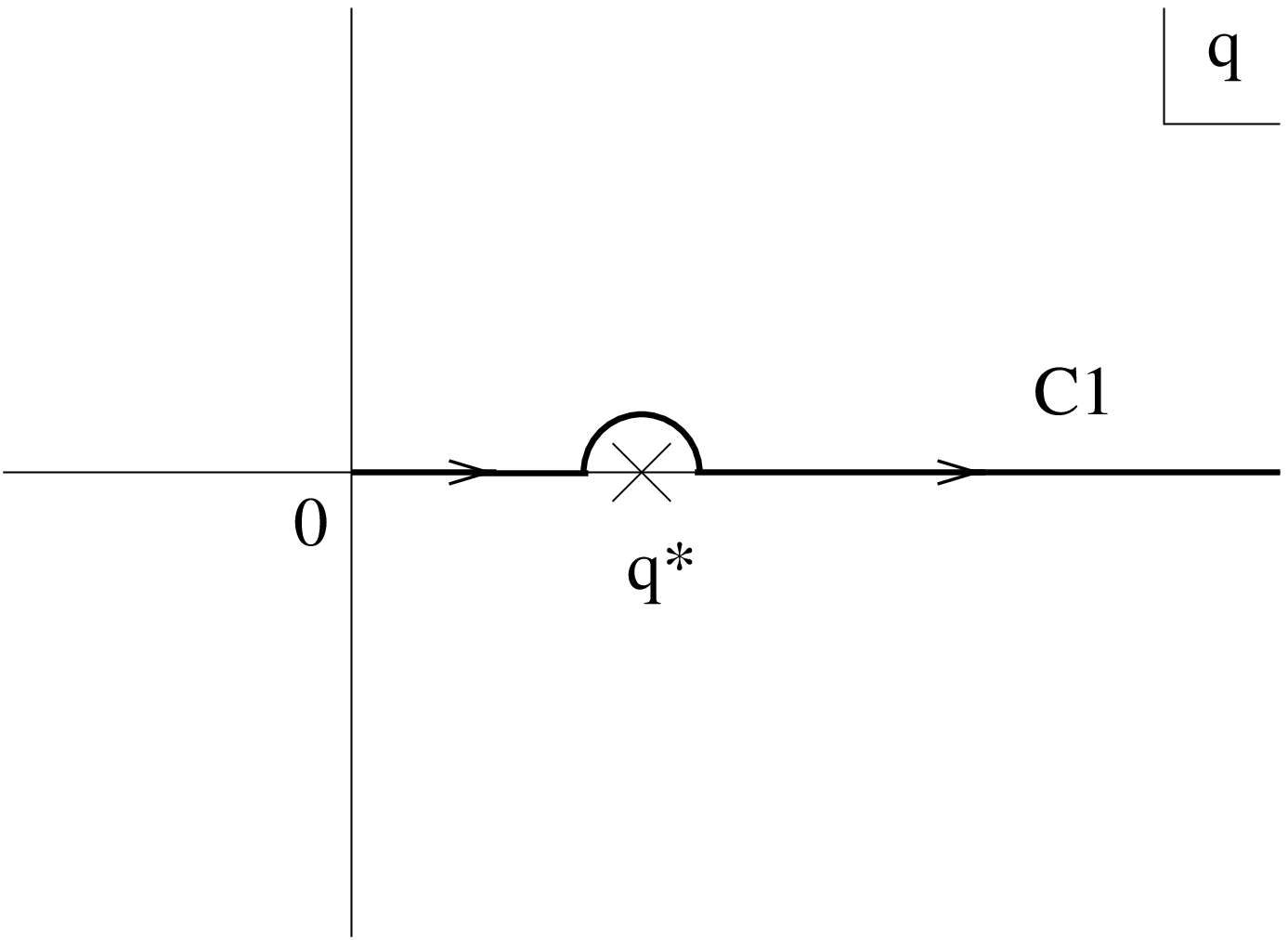} 
& \hspace{20mm}
\psfrag{C2}{$C_2$} 
\psfrag{q}{$q$} 
\psfrag{q*}{$q_*$} 
\psfrag{0}{\hspace{-1mm}\raise-1mm\hbox{$0$}}
\includegraphics[width=4.5cm]{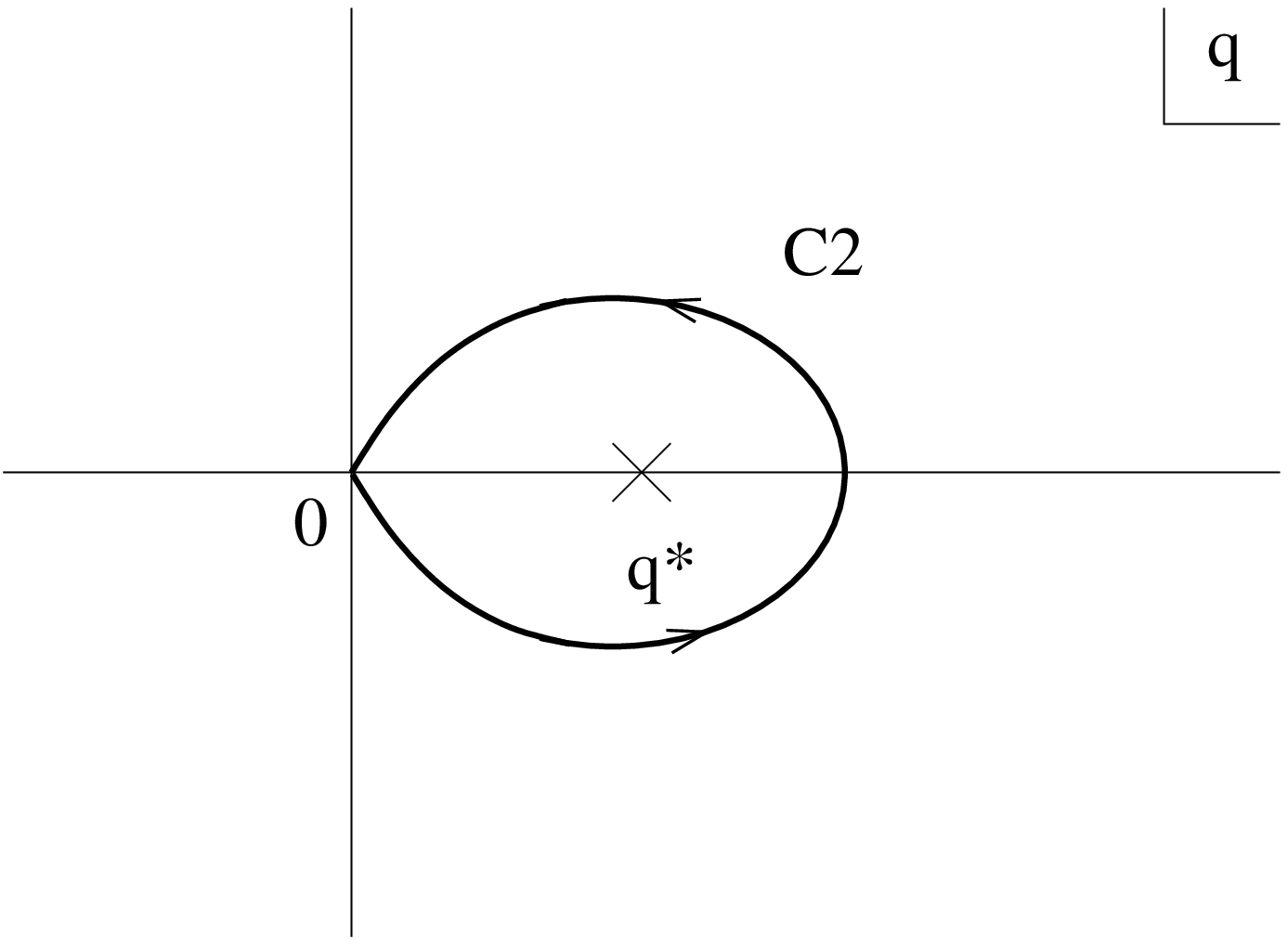}\\
(i) & \hspace{22mm}(ii)
\end{tabular}
\end{center}
\vspace*{-.5cm}
\caption{\small Integral contours $C_1$ and $C_2$ on the complex $q$-plane.
$q_*$ denotes the Landau singularity of $\alpha_S(q)$.
For 1-loop running, $q_*$ is a pole; for 2- and 3-loop running,
$q_*$ is a branch point.
In the latter case, branch cut is on the real axis starting from $q_*$
to $-\infty$.
\label{path}}
\end{figure}
The integral contours $C_1$ and $C_2$ are displayed 
in Figs.~\ref{path}(i),(ii),
respectively.\footnote{
We conjecture that the expressions (\ref{genform1})--(\ref{genform4})
are valid also beyond the 3-loop running, i.e. when the
higher $\beta_n$ and $a_n$ are incorporated,
as long as $\alpha_S=0$ remains to be the IR fixed point when 
$\alpha_S(q)$ is evolved along $C_1$.
}

The coefficient of the linear potential
can be expressed analytically for (a),(b),(c).
In the first two cases, the expressions read
\bea
&&
{\cal C}^{(a)}= \frac{2\pi C_F}{\beta_0} \,
\Bigl( \Lambda_{\overline{\rm MS}}^{\mbox{\scriptsize 1-loop}} \Bigr)^2 ,
\\ &&
{\cal C}^{(b)}= \frac{2\pi C_F}{\beta_0} \,
\Bigl( \Lambda_{\overline{\rm MS}}^{\mbox{\scriptsize 2-loop}} \Bigr)^2
\, \frac{e^{-\delta}}{\Gamma (1+\delta)} \,
\biggl[ 1 + \frac{a_1}{\beta_0} \, \delta^{-1-\delta} \, e^\delta \,
\gamma (1+\delta,\delta) \biggr] ,
\eea
where $\gamma (x,\tau) \equiv \int_0^\tau dt \, t^{x-1} \, e^{-t}$
represents the incomplete gamma function;
$\Lambda_{\overline{\rm MS}}^{\mbox{\scriptsize 1-loop}}$ and
$\Lambda_{\overline{\rm MS}}^{\mbox{\scriptsize 2-loop}}$
denote the Lambda parameters in the $\overline{\rm MS}$ scheme;
$\delta = \beta_1/\beta_0^2$.
In the case (c), ${\cal C}$ can be expressed in terms of confluent
hypergeometric functions except for the coefficient of $a_2$,
while the coefficient of $a_2$ can be expressed in terms of generalized
confluent hypergeometic functions.
Since, however, the expression is lengthy and not very illuminating,
we do not present it here.

The asymptotic behaviors of $V_C(r)$ for $r \to 0$ are same as those of
$V_{\rm QCD}(r)$ in the respective cases, as determined 
by RG equations.
The asymptotic behaviors of $V_C(r)$ for $r \to \infty$ are given
by $-4\pi C_F/(\beta_0 r)$ [the first term of eq.~(\ref{genform1})]
in all the cases.

As for ${\cal B}(N)$ and ${\cal D}(r,N)$, we have
not obtained simple expressions in the cases (b),(c), since 
analytic treatments are more difficult than in the case (a):
we have not
separated the divergent parts as $N \to \infty$ nor obtained the
asymptotic forms for $r \to 0$, $r \to \infty$.
Based on some analytic examinations, together with 
numerical examinations for $N \leq 30$, we conjecture
that ${\cal B}(N)$ and ${\cal D}(r,N)$ in the cases
(b),(c) have behaviors similar to those in the case (a).

\begin{figure}
\begin{center}
\hspace{-25mm}
\includegraphics[width=14cm]{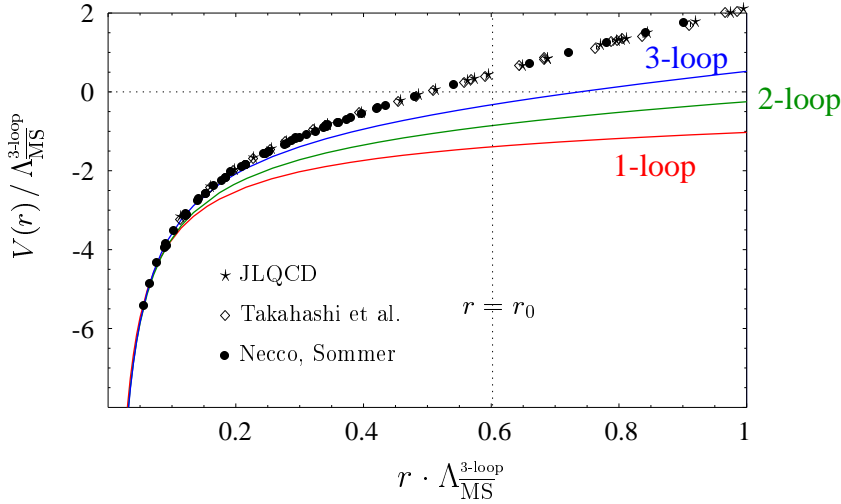}
\end{center}
\vspace*{-.5cm}
\caption{\small
Comparison of $V_C(r) + {\cal C} \, r$ corresponding to the cases (a),(b),(c)
(solid lines) and the lattice data \cite{latticedata}:
Takahashi et al.\ ($\diamond$), 
Necco/Sommer ($\bullet$), and JLQCD ($\star$).
\label{comp-lat}}
\end{figure}

Let us compare the ``Coulomb--plus--linear'' potential,
$V_C(r) + {\cal C} \, r$, for the three cases
when the number of quark flavors is zero.
We also compare them with
lattice calculations of the QCD potential in the quenched approximation.
See Fig.~\ref{comp-lat}.
We take the input parameter for 
$V_C(r) + {\cal C} \, r$ as $\alpha_S(Q)=0.2$, which corresponds to
$\Lambda_{\overline{\rm MS}}^{\mbox{\scriptsize 1-loop}}/Q =0.057$,
$\Lambda_{\overline{\rm MS}}^{\mbox{\scriptsize 2-loop}}/Q =0.13$, 
$\Lambda_{\overline{\rm MS}}^{\mbox{\scriptsize 3-loop}}/Q =0.12$.\footnote{
As well-known, when the strong coupling constant at some large
scale, e.g.\ $\alpha_S(m_b)$, is fixed, the values of
$\Lambda_{\overline{\rm MS}}^{\mbox{\scriptsize 1-loop}}$, 
$\Lambda_{\overline{\rm MS}}^{\mbox{\scriptsize 2-loop}}$, and
$\Lambda_{\overline{\rm MS}}^{\mbox{\scriptsize 3-loop}}$
differ substantially.
As a result, if we take a common value of $\Lambda_{\overline{\rm MS}}$ as the
input parameter, $V_C(r) + {\cal C} \, r$ for (a),(b),(c) differ
significantly at small distances,
where the predictions are supposed to be more accurate.
}
Then, the scale for each lattice data set is fixed using the
central value of the relation \cite{Capitani:1998mq}
$\Lambda_{\overline{\rm MS}}^{\mbox{\scriptsize 3-loop}}\, r_0=0.602(48)$,
where $r_0$ is the Sommer scale.
An arbitrary
$r$-independent constant has been added to each potential and 
each lattice data set to facilitate the comparison.
We see that $V_C(r) + {\cal C} \, r$ for (a),(b),(c) agree
well at small distances, whereas at large distances the potential
becomes steeper as $\alpha_S(q)$ accelerates in the IR region, i.e.\
${\cal C}^{(a)} < {\cal C}^{(b)} < {\cal C}^{(c)}$.
This feature is in accordance with the qualitative understanding 
within perturbative QCD \cite{bsv1,Sumino:2001eh,necco-sommer}.
The lattice data and $V_C(r) + {\cal C} \, r$ also agree well
at small distances,
while they deviate at larger distances.
More terms we include in $V_C(r) + {\cal C} \, r$, up to larger
distances the potential agrees with the lattice data.
If we increase the value of input $\alpha_S(Q)$,
$V_C(r) + {\cal C} \, r$ for (a),(b),(c) come closer to one another at
$r \cdot
\Lambda_{\overline{\rm MS}}^{\mbox{\scriptsize 3-loop}} > 0.1$.
[The relation between $V_C(r) + {\cal C} \, r$ for (c)
and the lattice data remains unchanged.]

\section{Discussion}
\label{s5}

In this section we discuss two issues:
non-uniqueness of the decomposition of $V_N(r)$ and
IR divergences of $a_n$ ($n \geq 3$).

How to decompose $V_N(r)$ for $N \gg 1$ into the 
$\{ r^{-1}, r^0, r^1, r^2 \}$ terms is not unique.
It is because $V_N(r)$ cannot be expanded in Laurent series
about $r=0$ or $r=\infty$ due to logarithmic corrections.
In fact, consider a function $f(r)$ which behaves as
$const. + K r$ for $r \ll r_1$ and which is ${\cal O}(1/r)$ for
$r \gg r_1$, 
where $r_1$ represents a typical scale inherent in $f(r)$,
e.g.\ $\displaystyle f(r) = - \frac{K r_1^2}{r + r_1}$;
then we may redefine
$\widetilde{V}_C(r) = V_C(r) + f(r)$,
$\widetilde{\cal C} = {\cal C} - K$,
$\widetilde{\cal D}(r,N) = {\cal D}(r,N) + K r - f(r) + const.$
as the Coulomb part, the coefficient of the linear part, and the
quadratic part, respectively.
In particular, this redefinition changes the coefficient of the
linear potential.

On the other hand, we may consider the decomposition 
(\ref{decgen})--(\ref{genform4}) to be an optimal decomposition
for $10 \leq N \leq 100$, on account of the following
consideration.
Suppose $K$ is of the same order of magnitude as ${\cal C}$.
In the case $r_1 \simlt \Lambda_{\overline{\rm MS}}^{-1}$,
since $V_C(r) + {\cal C} r$ is a good approximation of 
$V_N(r) \! - \! {\cal B}(N)$ for  
$r \simlt \Lambda_{\overline{\rm MS}}^{-1}$ (see Fig.~\ref{VN-CplusL}),
$\widetilde{V}_C(r) + \widetilde{\cal C} r$ cannot be a good approximation of
$V_N(r) \! - \! {\cal B}(N)$ for 
$r_1 \simlt r \simlt \Lambda_{\overline{\rm MS}}^{-1}$.
In the opposite case $r_1 > \Lambda_{\overline{\rm MS}}^{-1}$,
$\widetilde{V}_C(r)$ shows a linear-potential-like behavior
for $\Lambda_{\overline{\rm MS}}^{-1} \simlt r \simlt r_1$.
Then it is not very appropriate to regard $\widetilde{V}_C(r)$
as the ``Coulomb'' part.
\medbreak

As stated, $a_n$ for $n \geq 3$ contain IR divergences.
In the computation of the heavy quarkonium spectrum 
based on potential-NRQCD formalism \cite{pNRQCD}, 
IR divergences contained in $V_{\rm QCD}(r)$
are cancelled and the spectrum becomes finite at each order of the expansion 
in $\alpha_S$ \cite{kp,bpsv}.
Since IR divergences of $V_{\rm QCD}(r)$ originate from
the separation of ultrasoft scale in the computation of the spectrum,
it is natural to factorize the divergences from $V_{\rm QCD}(r)$ 
by introducing a factorization scale $\mu_f$
(IR cutoff).
In this case, $V_{\rm QCD}(r)$ is rendered finite as well as
dependent on  $\mu_f$.
The IR divergences can be absorbed into 
a non-local gluon condensate.
Thus, in the cases including $a_n$ for $n \geq 3$,
it is sensible to investigate the truncated series $V_N(r)$
corresponding to $V_{\rm QCD}(r)$ regularized in this way.
Another regularization scheme, which may be useful in comparison with
lattice computations, is the resummation of a certain class of diagrams
as done in \cite{adm}, which turns the IR divergences into
a finite contribution to the QCD potential.

At the present stage, it is unclear to which regularization scheme or
to which choice of $\mu_f$ in the factorization scheme 
the large-$\beta_0$ approximation correspond.
If we take the renormalon dominance picture rather strongly, we may
expect that the regularization scheme dependence of the
QCD potential is weak, and that the large-$\beta_0$
approximation makes sense quite generally.
The full computation of $a_3$ will bring the status clearer on this point.

One may consider that the leading ultrasoft logarithms of the
QCD potential \cite{USlog} are of the same order as the logarithms
resummed by $\beta_2$, so that they should be incorporated in
the case (c) of Sec.~\ref{s4}.
It is achieved by replacing the corresponding $V$-scheme coupling
constant as
\bea
\alpha_V(q) \to \alpha_V(q) + \frac{C_A^3}{6\beta_0} \, \alpha_S(q)^3 \,
\ln \biggl[ \frac{\alpha_S(q)}{\alpha_S(\mu_f)} \biggr] 
\eea
in (\ref{genform1})--(\ref{genform3}), where $C_A=3$ is the second
Casimir operator of the adjoint representation.
We have checked that this contribution is very small and scarcely changes
$V_C(r) + {\cal C} \, r$ for the case (c) displayed in Fig.~\ref{comp-lat}.
This feature is consistent with the analysis of \cite{Pineda}.

\section{Conclusions}
\label{s6}

We studied properties of the truncated perturbative series $V_N(r)$
of the QCD potential for $N \gg 1$;
the perturbative expansion of $V_{\rm QCD}(r)$ is truncated 
at ${\cal O}(\alpha_S^N)$
[$N = 6\pi/(\beta_0\alpha_S)$], where the term becomes
minimal according to the estimate based on the NLO renormalon.
$V_N(r)$ was examined in the large--$\beta_0$ approximation in Sec.~\ref{s3}.
We decomposed $V_N(r)$ into the
\{$r^{-1}$, $r^0$, $r^1$, $r^2$\} terms
(with logarithmic corrections in the $r^{-1}$ and $r^2$ terms)
and analyzed properties of each term.
The ``Coulomb'' and linear parts are finite as $N \to \infty$,
whereas the constant and quadratic parts diverge.
We argued that the quadratic part can be interpreted as representing
an ${\cal O}(\widetilde{\Lambda}^3 r^2)$ uncertainty.
For finite $N$, $V_N(r)$ is approximated well by
the sum of the ``Coulomb''
and linear parts (up to a constant) 
for $r \simlt \widetilde{\Lambda}^{-1}$ and $10 \leq N \leq 100$.
In Sec.~\ref{s4}, higher-order terms of $V_{\rm QCD}(r)$
were estimated using the RG analysis.
We decomposed $V_N(r)$ into four parts and studied 
properties of the ``Coulomb-plus-linear'' potential.
Analytic expressions for the linear potential are given in the
1-loop and 2-loop running cases.
As we incorporate 1-, 2- and 3-loop running of $\alpha_S(q)$,
the linear potential becomes steeper, as well as 
the ``Coulomb-plus-linear'' potential agrees with the lattice
data up to larger distances; cf Fig.~\ref{comp-lat}.
It is an interesting question whether the ``Coulomb-plus-linear'' potential
converges toward the lattice data
beyond 3-loop running.

The linear potential is proportional to
$\Lambda_{\overline{\rm MS}}^2\, r$ (as it should be unless it is zero,
since there is no other dimensionful parameter).
Naively one may think that such
a linear potential cannot be produced within perturbation theory, 
since the expansion of
$\Lambda_{\overline{\rm MS}}$ in $\alpha_S$ vanishes to all orders.
As eqs.~(\ref{VNbeta0}),(\ref{vrhoN}) show, however,
the linear potential is indeed inherent even in a finite-order 
perturbative expansion of $V_{\rm QCD}(r)$, due to dimensional transmutation.
In this regard, we note again that for certain finte $N$, 
$V_N(r)$ is approximated fairly well by the ``Coulomb-plus-linear'' potential
(Fig.~\ref{VN-CplusL}).

\section*{Acknowledgements}

The author is grateful to H.~Suganuma, M.~Tanabashi, and A.~Penin
for fruitful discussion.
He also thanks S.~Recksiegel for compiling the lattice data.

\end{document}